\documentclass[%
 reprint,
nofootinbib,
 amsmath,amssymb,
 aps
]{revtex4-1}
\pdfoutput=1
\usepackage{graphicx}
\usepackage[utf8]{inputenc}
\usepackage{flushend}
\usepackage{dcolumn}
\usepackage{bm}
\usepackage{balance}

\usepackage{caption}
\usepackage{subcaption}
\usepackage[normalem]{ulem}

\usepackage[colorlinks = true,
            linkcolor = blue,
            urlcolor  = blue,
            citecolor = blue,
            anchorcolor = blue]{hyperref}
\usepackage{verbatim}
\usepackage{color,ulem}
\usepackage[english]{babel}

\usepackage[utf8]{inputenc}
\input Starburst.fd
\newcommand*\initfamily{\usefont{U}{Starburst}{xl}{n}}\initfamily

\newcommand{\beq}{\begin{eqnarray}}
\newcommand{\eeq}{\end{eqnarray}}
\usepackage{amsmath}
\usepackage{tikz}
\usetikzlibrary{decorations.pathmorphing}
\usetikzlibrary{shapes.misc}
\tikzset{cross/.style={cross out, draw=black, minimum size=8*(#1-\pgflinewidth), inner sep=0pt, outer sep=0pt},
cross/.default={1pt}}
\usetikzlibrary{patterns,math}

\def\be{\begin{equation}}
\def\ee{\end{equation}}
\def\bea{\begin{eqnarray}}
\def\eea{\end{eqnarray}}

\begin{document}
\title{\large Holographic dissipative space-time supersolids}
\author{Peng Yang$^{1,2}$}
\author{Matteo Baggioli$^{2,3}$}\email{b.matteo@sjtu.edu.cn}
\thanks{These authors are listed alphabetically}
\author{Zi Cai$^{2,3}$}\email{zcai@sjtu.edu.cn}
\thanks{These authors are listed alphabetically}
\author{Yu Tian$^{1,4}$}\email{ytian@ucas.ac.cn}
\thanks{These authors are listed alphabetically}
\author{Hongbao Zhang$^5$}\email{hongbaozhang@bnu.edu.cn}
\thanks{These authors are listed alphabetically}

\affiliation{$^1$School of Physical Sciences, University of Chinese Academy of Sciences, Beijing 100049, China}
\affiliation{$^{2}$Wilczek Quantum Center, School of Physics and Astronomy, Shanghai Jiao Tong University, Shanghai 200240, China}
\affiliation{$^{3}$Shanghai Research Center for Quantum Sciences, Shanghai 201315, China}
\affiliation{$^4$Institute of Theoretical Physics, Chinese Academy of Sciences, Beijing 100190, China}
\affiliation{$^5$Department of Physics, Beijing Normal University, Beijing 100875, China}

\begin{abstract}  
Driving a system out of equilibrium enriches the paradigm of spontaneous symmetry breaking, which could then take place not only in space but also in time.  The interplay between temporal and spatial symmetries, as well as symmetries from other internal degrees of freedom, can give rise to novel nonequilibrium phases of matter. In this study, we investigate a driven-dissipative superfluid model using holographic methods and reveal the existence of a space-time supersolid (STS) phase which concomitantly breaks the time translation, spatial translation, and the internal U(1) symmetry. The holographic methods naturally include finite temperature effects, which enables us to explore the complex phase diagram of this model and observe a cascade of out-of-equilibrium phase transitions from the STS phase to a synchronized superfluid phase, and finally to a normal fluid phase, by increasing the temperature.
\end{abstract}

\maketitle

{\it Introduction}---The physics of quantum many-body systems out of equilibrium is much richer than its equilibrium counterpart, but less is known in general. As a prototypical example of a non-equilibrium quantum phase of matter, time crystals (TC) \cite{Wilczek2012}, which are characterized by spontaneous time translation symmetry breaking, have attracted considerable interests in various branches of modern physics \cite{Bruno2013,Watanabe2015,Sacha2015,Khemani2016,Else2016,Yao2017,Pizzi2021,Ye2021,McGinley2022}, including trapped ions \cite{Zhang2017},
nitrogen-vacancy center systems \cite{Choi2017},  quantum computation \cite{Mi2022,Frey2022}, and ultracold atoms \cite{Smits2018,Stehouwer2021,PhysRevLett.127.043602,Kongkhambut2022}. In realistic experimental setups, any quantum system is inevitably coupled to its surroundings, {\it i.e.}, to a thermal bath which inevitably induces dissipation. Understanding dissipative quantum time crystals is not only a question of practical experimental importance,  but also of fundamental significance due to its relevance to broader concepts such as the stability of non-equilibrium quantum matter against thermal fluctuations and the universality class of the associated non-equilibrium phase transitions. However, in spite of recent efforts made in classical TC systems \cite{Yao2020,Yue2022,Yue2023}, the effects of a thermal bath on a quantum time crystal are still poorly understood.

Recently, remarkable progress has been made in exploring non-equilibrium interacting quantum systems by exploiting the AdS/CFT correspondence (for a review, see Ref.\cite{Liu2019}). This holographic duality recasts the problem of strongly coupled quantum systems in the language of classical gravitational models in an asymptotically higher-dimensional Anti-de Sitter spacetime \cite{ZAANEN2015,Hartnoll:2016apf}.  More importantly, in this framework, the existence of a black hole in the gravitational background is equivalent to the presence of a thermal bath for the boundary quantum system, hence it enables us to naturally include finite temperature effects without the need of uncontrolled phenomenological modeling. In the last decade, the holographic duality has provided new opportunities to explore the dissipative dynamics of out-of-equilibrium quantum many-body systems, including non-equilibrium steady states \cite{PhysRevLett.109.091601,Bhaseen2015,Kundu2019}, out-of-equilibrium phase transitions \cite{Nakamura:2012ae,Guo:2018mip,Imaizumi:2019byu}, quantum quenches \cite{Bhaseen2013,Chesler2015}, driven systems \cite{Li:2013fhw,Auzzi:2013pca,Rangamani:2015sha,Baggioli:2021tzr} and quantum turbulence \cite{doi:10.1126/science.1233529,Adams:2013vsa,Lan:2016cgl}. In addition to that, AdS/CFT has been very successful in describing strongly-correlated phases of matter exhibiting spontaneous symmetry breaking of both internal (\textit{e.g.}, superfluids \cite{Hartnoll2008}) and spatial translation (\textit{e.g.}, solids \cite{Alberte:2017oqx}, charge density waves \cite{Baggioli:2022pyb}, and supersolids \cite{Baggioli:2022aft,Yang:2023vxz}) symmetries, but has never been applied so far to the case of time translations.

\begin{figure}[htb]
\includegraphics[width=0.9\linewidth]{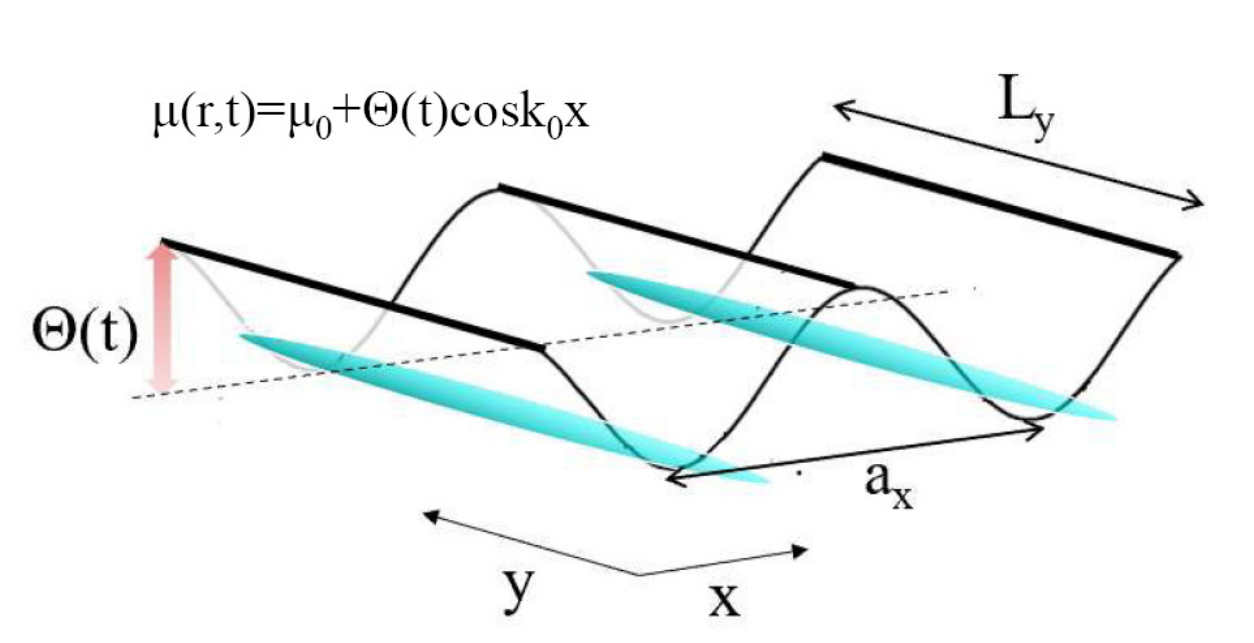}
\caption{(Color online) Sketch of the periodic and inhomogeneous chemical potential applied to the superfluid. $\Theta(t)$ is a periodic function with frequency $\omega_d$ and wavelength $a_x=2\pi/k_0$.} \label{fig:fig1}
\end{figure}

In this study, we investigate the late-time dynamics of a periodically-driven holographic superfluid, and we uncover the existence of a space-time crystal phase which simultaneously breaks the continuous spatial translation symmetry and the discrete time translation symmetry (DTTS). This space-time crystal also breaks the U(1) symmetry as a superfluid and it can thus be considered as a space-time supersolid (STS). The emergent space-time orders appear to be intertwined into a rich phase diagram which can be explored via a linear instability analysis based on Floquet theory. The holographic methods allow to further study the effect of thermal fluctuation on this quantum STS. It is observed that the system undergoes two continuous out-of-equilibrium phase transitions, each of which restores the DTTS and U(1) symmetry respectively. In particular, the system goes from the STS phase to the synchronized phase when the temperature increases, in accordance with the physical intuition that crystals melt upon heating (similar to classical TC \cite{Yue2022}).\\

{\it Methods}---A holographic superfluid \cite{Hartnoll2008} is described by an Abelian-Higgs model in a (3+1)-dimensional AdS black hole spacetime:
\begin{equation}
S=\int d^4 x \sqrt{-g} \left({R+6}-\frac{F^2}4-|D\Psi|^2-m^2|\Psi|^2 \right) \label{eq:AH}
\end{equation}
where $\Psi$ is a complex scalar field and $A_\mu$ is a U(1) gauge field with
$F_{\mu\nu}=\partial_\mu A_\nu-\partial_\nu A_\mu$ and $D=\partial_\mu -iA_\mu$. For simplicity, we have set the AdS radius to unit. Throughout the paper, we work in the probe limit in which the backreaction of all the matter fields on the metric is assumed to be negligible, which is a justified assumption whenever the temperature is not too low \cite{HHH2008}. Hence, the equations of motion (EOM) for the scalar and gauge fields read:
\begin{align}
&\left(D^\mu D_{\mu}-m^2\right)\Psi=0 \\
&\nabla_\mu F^{\mu\nu}=J^\nu_\Psi=i(\Psi^* D^\nu \Psi-\Psi(D^\nu \Psi)^*).
\end{align}
The $(3+1)$-dimensional background spacetime is a Schwarzschild-AdS black hole:
\begin{equation}
ds^2=\frac {1}{z^2}\left(-f(z)dt^2+\frac{1}{f(z)}dz^2+ dx^2+dy^2\right).
\end{equation}
The holographic coordinate $z$ spans from the AdS boundary $z=0$ to the location of the horizon $z=z_H$. The blackening factor is given by $f(z)=1-\frac{z^3}{z_H^3}$. The temperature of the dual field theory is $T_H=\frac {3}{4\pi z_H}$. In the rest of the manuscript, we set $m^2=-2$, and we further choose the axial gauge $A_z=0$.

Following the AdS/CFT dictionary \cite{Hartnoll:2016apf}, the dual field theory is a large $N$ CFT with a global U(1) symmetry and a corresponding conserved current $J_\mu$. The gauge field $A_\mu$ is dual to the current operator $J_\mu$, and the complex scalar $\Psi$ is dual to a scalar operator $\mathcal{O}$ with conformal dimension $\Delta=2$, charged under the U(1) global symmetry. Close to the AdS boundary ($z\rightarrow 0$), the matter fields have the following time-dependent asymptotic expansion:
\begin{eqnarray}
\Psi&=& z\,\Psi_1(t,\mathbf{r})+z^2 \Psi_2(t,\mathbf{r}),\\
A_t&=&\mu(t,\mathbf{r})-z\,\rho(t,\mathbf{r}), \label{eq:mu}
\end{eqnarray}
where $\Psi_1$ is the source for the operator $\mathcal{O}$, while $\mu$ and $\rho$ are the chemical potential and the conserved charge density respectively. With $\Psi_1=0$, the condensate $\langle\mathcal{O}\rangle=\Psi_2\neq 0$ forms below a certain critical temperature, signaling the onset of U(1) spontaneous symmetry breaking to the superfluid phase from the normal fluid phase. In the following, we consider a periodic chemical potential (see Fig.\ref{fig:fig1}):
\begin{equation}
    \mu(\mathbf{r},t)=\mu_0+\Theta(t)\cos2\pi x/a_x
\end{equation}
which breaks translation invariance along the $x$ direction, as in \cite{Yang:2021ssm}. The amplitude $\Theta(t)$ is modulated in time, with a period $T=2\pi/\omega_d$,
\begin{equation}
    \Theta(t)=\Theta_0+\Theta\sin\omega_d t.
\end{equation}
Periodic boundary condition in the $x,y$ directions are chosen using a box $L_x \times L_y$. More details about the numerical methods can be found in the supplementary material (SM).

Without the periodic potential, the critical temperature for the onset of superfluidity is given by $T_c= 0.0587 \mu_0$. For later convenience, we define a dimensionless reduced temperature $\tau=T_H/T_c$.

\begin{figure}[htb]
\includegraphics[width=0.99\linewidth,bb=35 74 790 750]{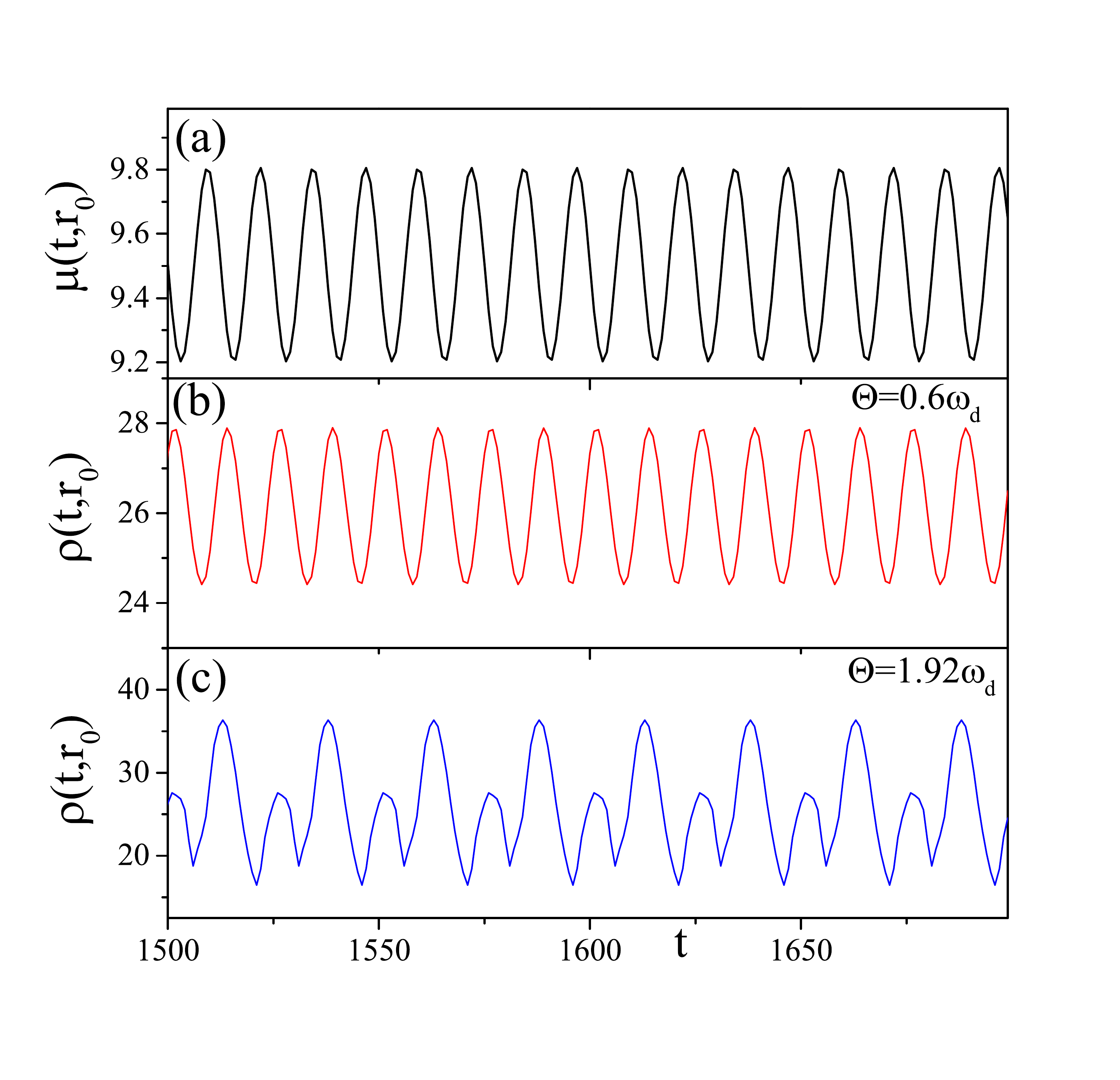}
\caption{(Color online) (a) The time-dependent chemical potential driving the system out of equilibrium. The late-time dynamics of the local density $\rho(t,\mathbf{r}_0)$ at $\mathbf{r}_0=(0,0)$ in (b) the synchronized phase with weak driving, and (c) the space-time crystal phase with strong driving. Parameters are chosen as $\mu_0=5.4$, $\Theta_0=4.1$ and $\omega_d=0.16\pi$.}
	\label{fig:fig2}
	\end{figure}

\begin{figure}[htb]
\includegraphics[width=0.99\linewidth,bb=150 60 940 550]{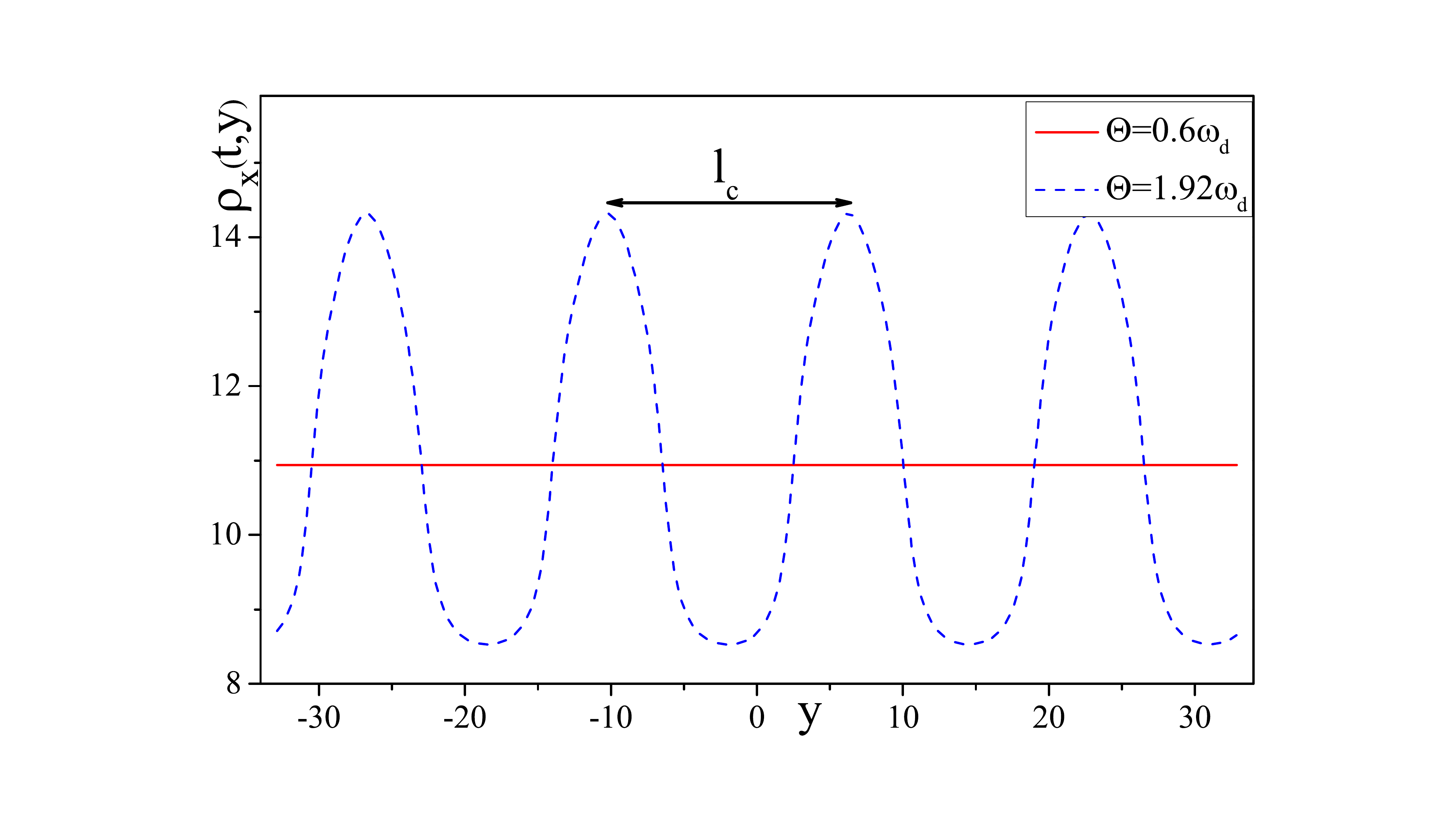}
\caption{(Color online) Spatial profile of the normalized average density $\rho_x(t,y)=\frac 1{L_x}\int dx \rho(t,\mathbf{r})$ along the y-direction in the synchronized phase with weak driving (red solid), and in the space-time crystal phase with strong driving (dashed blue). }
	\label{fig:fig3}
	\end{figure}

{\it Synchronized phase versus space-time crystal}---To characterize the crystalline pattern, we focus on the charge density $\rho(t,\mathbf{r})$, which can be extracted from Eq.(\ref{eq:mu}). In particular, it is sufficient for us to check the late-time dynamics of $\rho(t,\mathbf{r}_0)$ at a specific position $\mathbf{r}_0=(0,0)$.  Fig.\ref{fig:fig2} shows that, in the presence of a weak driving, $\rho(t,\mathbf{r}_0)$ oscillates with a period identical to that of the external driving, thus the system is in a synchronized phase. However, when the driving amplitude exceeds a critical value, the system displays a period-doubling discrete time crystal phase, which spontaneously breaks the $Z_2$ translational symmetry of the system, $\rho(t,\mathbf{r}_0)=\rho(t+2T,\mathbf{r}_0)\neq\rho(t+T,\mathbf{r}_0)$. This period doubling phenomenon is reminiscent of the Faraday waves observed in Bose-Einstein condensate systems \cite{Engels2007}, which can be classically described using the Gross-Pitaevskii (GP) equation \cite{Nicolin2007}. 

 Next we study the spatial profile of these two non-equilibrium phases. The translation invariance along the $x$-direction is broken by the inhomogeneous chemical potential, while the system is translation invariant in the $y$-direction. To characterize the spontaneous symmetry breaking of translations along the $y$-direction, we define a normalized average density $\rho_x(t,y)=\frac 1{L_x}\int dx \rho(t,\mathbf{r})$. As shown in Fig.\ref{fig:fig3}, at a fixed time slice $t=t^*$, the normalized density is homogeneous along the $y$-direction in the synchronized phase, thus it retains the continuous translation symmetry of the system. On the contrary, in the presence of a strong driving, the homogeneous pattern is no longer stable, and a crystalline phase with a characteristic length $l_c$ spontaneously emerges. The continuous translation symmetry in the $y$-direction is spontaneously broken into a discrete one, $\rho_x(t^*,y)=\rho_x(t^*,y+l_c)$. In a summary, a strong periodical driving in the non-equilibrium holographic superfluid leads to a space-time crystal phase, which simultaneously breaks the spatial and temporal translation symmetry, a STS.\\

\begin{figure}[htb]
\includegraphics[width=0.99\linewidth,bb=50 43 970 540]{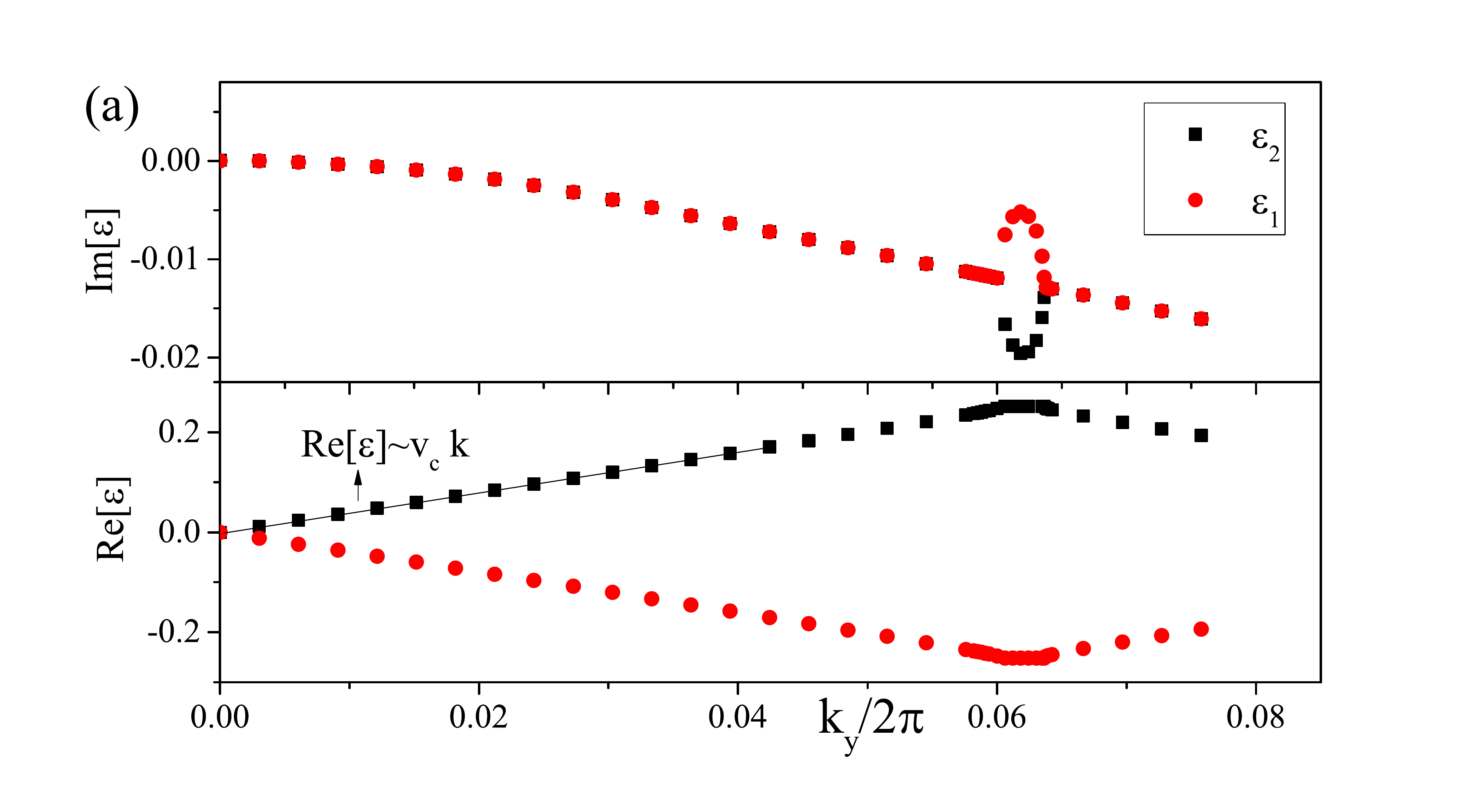}
\includegraphics[width=0.99\linewidth,bb=50 43 970 540]{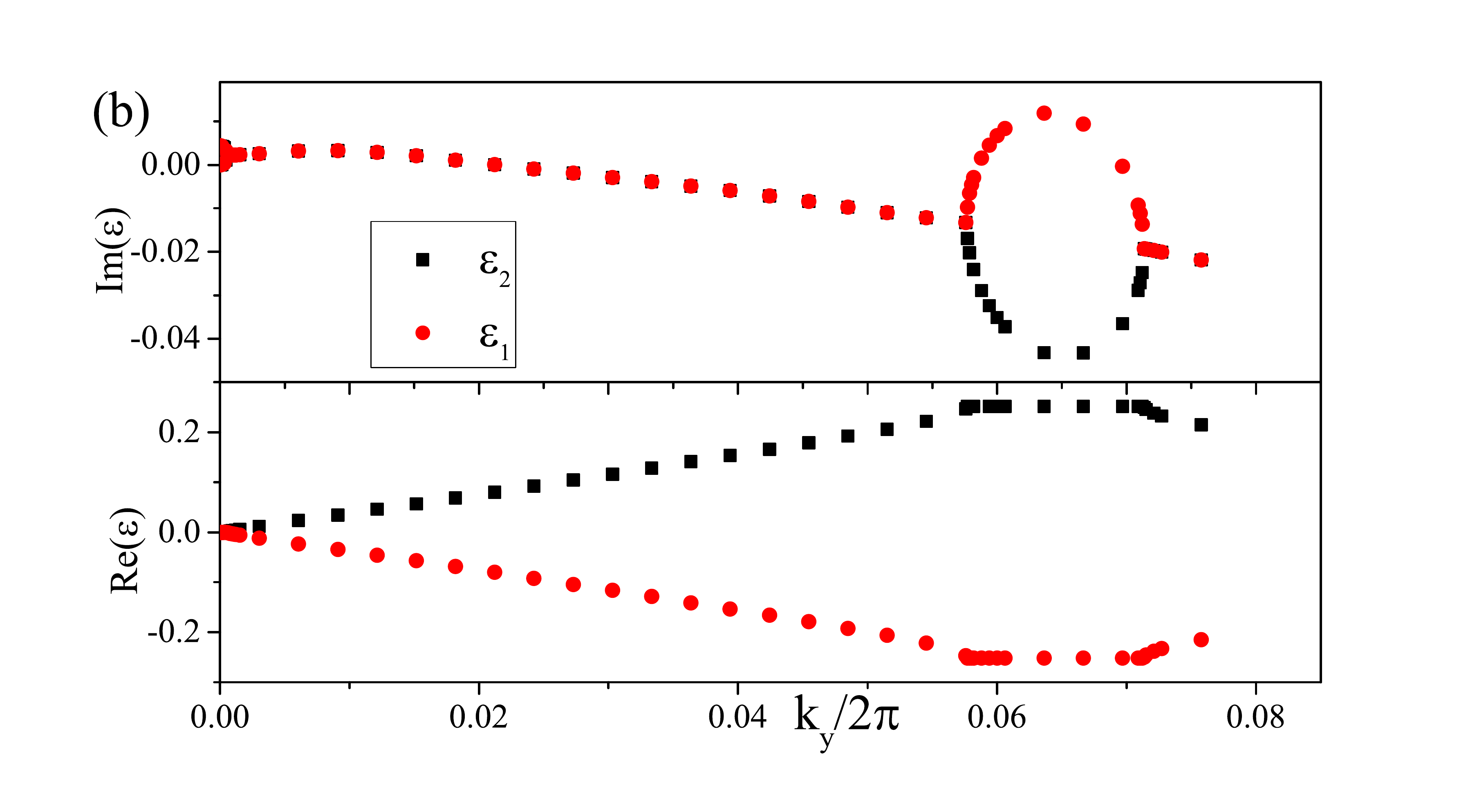}
\caption{(Color online) The $k_y-$dependence of the imaginary (upper panel) and the real parts (lower panel) of the $1st$ and $2nd$ eigenvalues of the Floquet matrix $\mathbb{H}_{k_y}$ in the linear stability analysis for (a) the synchronized phase with $\Theta=0.6\omega_d$ and (b) the space-time crystal phase $\Theta=1.92\omega_d$. The dashed line in (a) indicates the dispersion of the Goldstone modes for the synchronized SF $\Re[\epsilon]\sim v_c k_y$ with sound speed $v_c=1.21\pi$, as computed from hydrodynamics.}
\label{fig:fig4}
\end{figure}

{\it Linear instability analysis}---The instability of the synchronized phase and the nature of the holographic space-time crystal can be understood in the framework of linear instability analysis. We start from a synchronized solution where both $\Psi^s$ and $A^s_\mu$ periodically oscillates in time with a period $T$, and are homogeneous along the $y$-direction. To study the stability of this solution, we introduce the following perturbations:
\begin{eqnarray}
\Psi(t,z,x,y)&=&\Psi^s(t,z,x)+e^{ik_y y}\delta \Psi(t,z,x),  \label{eq:dpsi}\\
A_\mu(t,z,x,y)&=&A^s_\mu(t,z,x)+e^{ik_y y}\delta A_\mu(t,z,x), \label{eq:dAmu}
\end{eqnarray}
where $k_y=\frac {2\pi n}{L_y}$ with $n$ an integer. By substituting Eq.\eqref{eq:dpsi} and \eqref{eq:dAmu} into the EOM, and keeping only the linear terms in $\delta\Psi$ and $\delta A_\mu$, one obtains the EOM for the perturbations $\vec{\delta}_{k_y}=[\delta\Psi,\delta A_\mu]^T$ which can be written in vectorial form as
\begin{equation}
\frac{d \vec{\delta}_{k_y}}{dt} = \mathbb{M}_{k_y}(t)\vec{\delta}_{k_y}. \label{eq:EOMk}
\end{equation}
with the matrix $\mathbb{M}_{k_y}(t)=\mathbb{M}_{k_y}(t+T)$. The periodicity of $\mathbb{M}_{k_y}(t)$ enables us to employ the Floquet description for the stroboscopic dynamics in Eq.\eqref{eq:EOMk} and derive a time-independent Floquet matrix $\mathbb{H}_{k_y}$ satisfying:
\begin{equation}
\mathbb{U}_{k_y}=e^{-i\mathbb{H}_{k_y} T}=\mathcal{T}  e^{\int_0^T dt \mathbb{M}_{k_y}(t)},\label{eq:Floquet}
\end{equation}
where $\mathcal{T}$ is the time-ordering operator $\mathcal{T}  e^{\int_0^T dt \mathbb{M}_{k_y}(t)}=e^{dt \mathbb{M}_{k_y}(T-dt)}\cdots e^{dt \mathbb{M}_{k_y}(dt)} e^{dt \mathbb{M}_{k_y}(0)}$, and $\mathbb{U}_{k_y}$ is the evolution operator within one period, $\vec\delta_{k_y}(T)=\mathbb{U}_{k_y}\vec\delta_{k_y}(0)$.

The stability of the background solution $\Psi^s$ and $A^s_\mu$ depends on the imaginary part of the eigenvalues of the Floquet matrix $\mathbb{H}_{k_y}$. For our purpose, it is sufficient to focus on the eigenvalues with the largest and second largest imaginary parts, denoted respectively by $\varepsilon _{k_y}^1$ and $\varepsilon _{k_y}^2$. In particular, whether the background solution is stable or unstable depends on whether the imaginary part of $\varepsilon_{k_y}^1$ is greater than or less than zero, where its real part $\Re[\varepsilon_{k_y}^1]$ indicates the frequency of the oscillation accompanying the exponential divergence or decay.

\begin{figure}[htb]
\includegraphics[width=0.99\linewidth,bb=36 53 950 550]{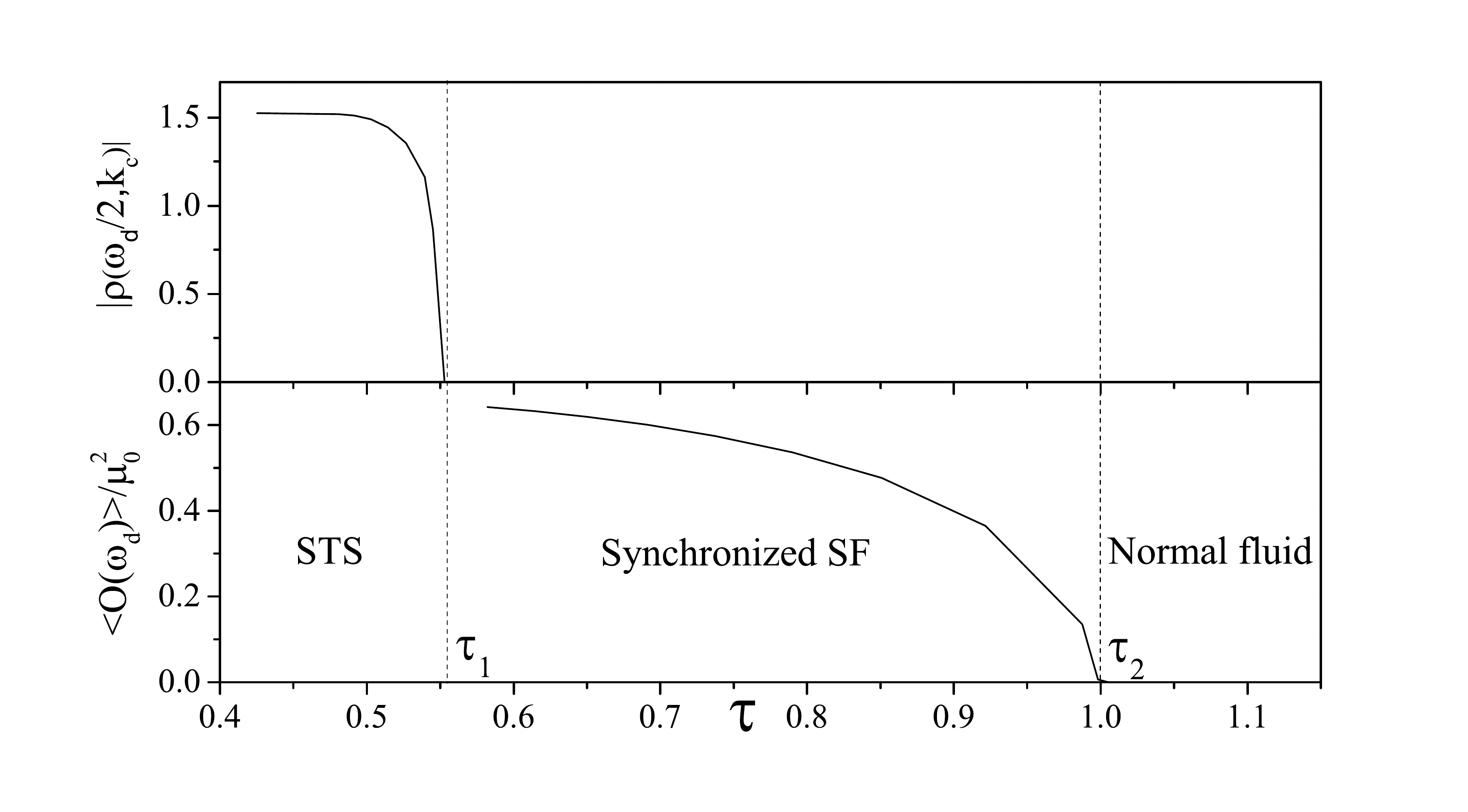}
\caption{(Color online) The dependence of the STS order parameter  (top panel) and the synchronized SF order parameter  (bottom panel) as a function of the reduced temperature $\tau$.} \label{fig:fig5}
\end{figure}

The real and imaginary parts of $\varepsilon_{k_y}^1$ and $\varepsilon_{k_y}^2$ as a function of $k_y$ are plotted in Fig.\ref{fig:fig4} with $\Re[\varepsilon_{k_y}^1]=-\Re[\varepsilon_{k_y}^2]$. The corresponding low-energy excitations are the left and right propagating superfluid sound modes which arise because of the coupling of the Goldstone mode with charge density fluctuations. This is confirmed by comparing the data for the real part of the dispersion at low wave-vectors with the expectations from relativistic superfluid hydrodynamics \cite{Amado:2009ts,Arean:2021tks}, $v_c^2=\rho_s/(\mu \chi)$ ($\rho_s$ and $\chi$ being respectively the superfluid density and the charge susceptibility). In both weak and strong drivings, we find two flat bands appear in some intermediate $k_y$ regime, where $\Re[\varepsilon_{k_y}^1]=-\Re[\varepsilon_{k_y}^2]=\frac{\omega_d}2$  is accompanied by the degeneracy lift of $\Im[\varepsilon_{k_c}^1]$ and $\Im[\varepsilon_{k_c}^2]$. With the observation that the two modes colliding right at the edge of Floquet zone give rise to the degeneracy of their real parts, the above pattern can accordingly be well understood by the similar argument arising in the black hole dynamics \cite{QNM2016}, Bose-Einstein condensates \cite{BEC2005,BEC2008}, and other systems \cite{Weinberg1940, Fulling1976}. In the presence of weak driving, $\Im[\varepsilon_{k_y}^1]$ is always negative for arbitrary $k_y$ as shown in Fig.\ref{fig:fig4} (a), indicating that the homogeneous synchronized phase is stable against perturbations. 
On the contrary, in the strongly driven case, $\Im[\varepsilon_{k_y}^1]$ becomes positive and reaches its maximum at $k_y=k_y^c$. These results indicate that the homogeneous synchronized phase is not stable against the spatial fluctuations along the $y$-direction. In other words, the system will spontaneously develop a crystalline pattern with a characteristic length $l_c=\frac{2\pi}{k_c}$. Furthermore, a non vanishing $\Re[\varepsilon_{k_y}^1]=\frac{\omega_d}2$ at $k_y=k_c$ indicates that the spatial crystal is accompanied by a temporal oscillation with a frequency $\omega=\frac{\omega_d}2$, which explains the period doubling in the space-time crystal observed in the real-time numerical simulations in Fig.\ref{fig:fig2}.

{\it Non-equilibrium phase transitions and phase diagram}---Now we consider the effects of temperature. In the presence of a strong driving, at low temperature, the system is in a STS phase, which is characterized by Bragg peaks in the Fourier spectrum for the density distribution $\rho(\omega,k_y)=\int dt dy e^{-ik_y y+i\omega t}\rho_x(t,y)$ located at ($\pm k_c,\pm\frac{\omega_d}2$). As temperature increases, the height of the peak $|\rho(\frac \omega2,k_c)|$ decreases and finally vanishes at a critical temperature $\tau_1$, which indicates a phase transition from a STS to a synchronized superfluid (SF) phase. In the intermediate temperature regime, the crystalline order is suppressed by thermal fluctuations, while the SF order survives. This synchronized SF phase is characterized by a nonzero SF order parameter $\langle\mathcal{O}(t,\mathbf{r})\rangle$. We consider its Fourier component $\mathcal{O}(\omega_d)=\int dt  e^{i\omega_d t}\mathcal{O}(t,\mathbf{r}_0)$. As shown in Fig.\ref{fig:fig5}, $|\mathcal{O}(\omega_d)|$ vanishes at a temperature $\tau_2$, indicating a phase transition from a synchronized SF to a normal fluid. 

\begin{figure}[htb]
\includegraphics[width=0.99\linewidth,bb=16 65 750 530]{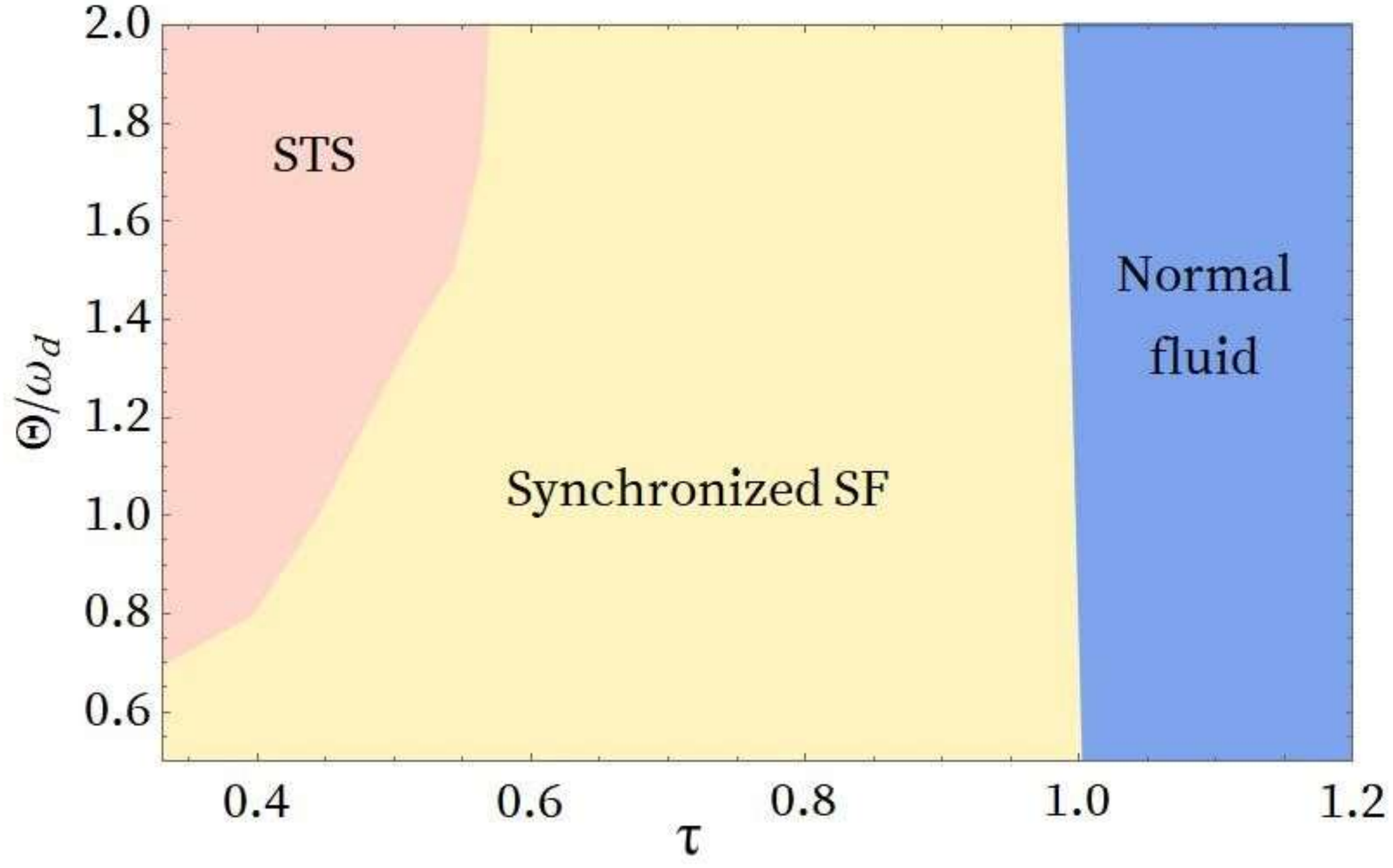}
\caption{(Color online) Phase diagram in terms of the driving amplitude $\Theta$ and the reduced temperature $\tau$.} \label{fig:fig6}
\end{figure}

The phase diagram in terms of the driving amplitude $\Theta$ and temperature $\tau$ is plotted in Fig.\ref{fig:fig6}, which shows that the holographic STS exists in the regime with strong driving and low temperature. As the temperature increases, the system will experience two continuous non-equilibrium phase transitions, which are characterized by the restoration of the DTTS and U(1) symmetry respectively. The relationship between the DTTS and U(1) symmetry breaking was discussed in Ref.\cite{Dai2022}. In general, the $Z_2$ long-range temporal crystalline order is unstable against the presence of stochastic $\pi$-phase shift in time domain (for the same reason that 1D spatial $Z_2$ long-range order is unstable against the propagation of the kink excitations activated by thermal fluctuations). However, for a system that simultaneously breaks the DTTS and U(1) symmetry, a $\pi$-phase shift in time domain is accompanied by a phase slip $\phi\rightarrow \phi+\pi$ in the U(1) symmetry breaking order parameter, which is energetically disfavored in the U(1) symmetry breaking phase. As a consequence, the discrete time crystalline order is protected by the U(1) symmetry breaking, thus its critical temperature cannot be larger than that of the SF condensate, \textit{i.e.} $\tau_2>\tau_1$, as observed in Fig.\ref{fig:fig6}.\\

{\it Conclusions}---In summary, by using holographic methods, we studied the late-time out of equilibrium dynamics of a driven-dissipative quantum system, and discovered a space-time supersolid phase which simultaneously breaks the spatial and temporal translation symmetries together with the internal U(1) symmetry. As the temperature increases, we find that the thermal fluctuations first restore the DTTS, and then the U(1) symmetry, leaving an intermediate synchronized SF phase between them. This cascade of non-equilibrium phase transitions from STS to normal fluid could be checked in future experiments, for example using ultra-cold atom systems in an optical lattice. Further developments of our analysis will include the generalizations of our results
to time crystal phases with different translation
symmetry breaking, for instance, the continuous time crystal \cite{Kongkhambut2022} and time quasicrystal \cite{Autti2018}, where the interplay between the spatial U(1) symmetry and continuous time translation symmetry might give rise to novel non-equilibrium phases and phase transitions.

From a physical point of view, we emphasize that the probe limit we are working with corresponds to considering the dissipative time crystal in open quantum systems (see \cite{doi:10.1146/annurev-conmatphys-031119-050658,RevModPhys.95.031001} for a classification and discussion about the various scenarios), where the number of degrees of freedom (dof) of the thermal bath is parametrically large. This is a common limit for open quantum systems in quantum optics, atomic and molecular physics where one assumes that the backreaction of the system on the bath remains negligible. From the gravitational point of view, this argument can be made explicit in probe brane setups, where the suppression is controlled by $N_f/N_c$, with $N_f,N_c$ respectively the number of colors and flavors (see for example \cite{Karch:2008uy}). 
This hierarchy allows to dissipate heat very efficiently and maintain a steady state for a long timescale without destroying the time-crystalline order, as realized experimentally in \cite{PhysRevLett.127.043602,Kongkhambut2022}. Furthermore, as shown explicitly for a scalar toy model in the SM, even with backreaction, the heating induced by the driving can be kept parameterically small by controlling the relative number of dof between the bath and the system. This implies that the time-crystalline order survives in the backreaction limit up to a time-scale which can be made arbitrarily long.\\

{\it Acknowledgments}---We would like to thank Hong Liu for his valuable suggestions regarding the viability of the probe limit as well as the heating issue. ZC is supported by the National Key Research and Development Program of China (Grant No.2020YFA0309000), Natural Science Foundation of China (Grant No.12174251), Natural Science Foundation of Shanghai (Grant No.22ZR142830). MB acknowledges the support of the Shanghai Municipal Science and Technology Major Project (Grant No.2019SHZDZX01) and the sponsorship from the Yangyang Development Fund. YT is partly supported by the Natural Science Foundation of China (Grant Nos.11975235 and 12035016).  HZ is partly supported by the National Key Research and Development Program of China (Grant No.2021YFC2203001) and Natural Science Foundation of China (Grant No.12075026).

\bibliographystyle{apsrev4-1}
\bibliography{holo}

\onecolumngrid
\appendix 
\clearpage
\renewcommand\thefigure{S\arabic{figure}}    
\setcounter{figure}{0} 
\renewcommand{\theequation}{S\arabic{equation}}
\setcounter{equation}{0}
\renewcommand{\thesubsection}{SM\arabic{subsection}}
\section*{Supplementary Material}
In this supplementary Material, we provide more details about the holographic computations, the numerical methods and further evidence for the existence of a time crystal phase. {We also present a holographic toy model with backreaction to prove that the heating rate induced by the driving can be controlled using the backreaction strength, leading to pre-thermalized phases of matter.}

\subsection{Methods}\label{s1}

To numerically simulate the dynamics of holographic superfluid system under external driving, one needs to numerically solve the equations of motions for the bulk fields $\Psi$ and $A_\mu$ which, in Eddington-Finkelstein coordinates, are given by
\begin{eqnarray}
\partial_t \partial_z \psi&=&\partial_z\left(\frac{f(z)}{2}\partial_z\psi \right)+\frac{1}{2}\partial^2\psi-i\textbf{A}
\cdot\partial\psi +iA_t\partial_z\psi 
-\frac{i}{2}\left(\partial\cdot\textbf{A}-\partial_zA_t\right)\psi
-\frac{1}{2}\left(z+\textbf{A}^2\right)\psi, \label{dphi}\\
\partial_t \partial_z \textbf{A}&=&\partial_z\left(\frac{f(z)}{2}\partial_z\textbf{A}\right)-|\psi|^2\textbf{A}+\textrm{Im}(\psi^*\partial\psi)
+\frac{1}{2}\left[\partial\partial_zA_t+\partial^2\textbf{A} -\partial\partial \cdot\textbf{A}\right], \label{eqeom} \\
\partial_t\partial_z A_t&=&\partial^2A_t-\partial_t\partial\cdot\textbf{A}+f(z)\partial_z\partial\cdot\textbf{A} -2A_t|\psi|^2
+2\textrm{Im}(\psi^*\partial_t\psi)-2f(z)\textrm{Im}(\psi^*\partial_z\psi),
\label{eqAt}
\end{eqnarray}
together with the constraint
\begin{equation}
0=-\partial_z^2 A_t+\partial_z\partial\cdot\textbf{A}-2\textrm{Im}(\psi^*\partial_z\psi).\label{constrain}
\end{equation}
In the above equations, we have defined $\psi=\Psi/z$ and $\textbf{A}=(A_x, A_y)$. The metric is given by
\begin{equation}
    ds^{2}=\frac{1}{z^{2}}\left(-f(z) d t^{2}- dtdz+dx^{2}+dy^{2}\right)\,.
\end{equation}

When performing the full non-linear simulation, one considers the bulk fields as a function of time $t$ and space coordinates $z,x,y$. In the time direction, we adopt a fourth order Runge-Kutta method with time step $\Delta t=0.01$. In the spatial directions, the periodic box is taken as $15 \times 66$ and the horizon is set at $z_h=1$ using the scaling symmetries of the system. We use pseudo-spectral methods with the number of grid points taken as $25\times 65 \times 255$.

We impose the following boundary conditions for the bulk fields
\begin{equation}
\psi|_{z=0}=0, \ \ A_t|_{z=0}=\mu, \ \ \textbf{A}|_{z=0}=0.\label{bcs}
\end{equation}
so that the evolution of $\psi$ and $\textbf{A}$ are governed by Eq.\eqref{dphi} and Eq.\eqref{eqeom}, respectively. The above boundary conditions correspond to set the source for the dual scalar operator to zero and the U(1) boundary current to zero as well, while allowing for the presence of a finite chemical potential $\mu$. The behavior of $A_t$ can be obtained from the constraint equation with one additional boundary condition imposed at AdS boundary
\begin{equation}
\partial_z A_t|_{z=0}=-\rho\,.
\end{equation}
Here, the particle number density $\rho$ satisfies the current conservation law, $\partial_\mu J^\mu=0$, which is equivalent to assuming Eq.\eqref{eqAt} on the boundary $z=0$.

To perform the linear instability analysis, one first turns the dependence of the bulk field on the $y$ direction off. Then, $A_y$ can be consistently set to zero. With the same boundary conditions used in non-linear evolution, we drive this reduced $1+2$ dimension system into a steady Floquet state $\psi^{s}$ and $A_{\mu}^{s}$, in which the system is oscillating in synchronization with the external driving. 

Having the background state at hand, we introduce the following perturbations 
\begin{eqnarray}
\psi_{R,I}(t,z,x,y)&=&\psi_{R,I}^s(t,z,x)+e^{ik y}\delta \psi_{R,I}(t,z,x),  \label{peq:dpsi}\\
A_\mu(t,z,x,y)&=&A^s_\mu(t,z,x)+e^{ik y}\delta A_\mu(t,z,x), \label{peq:dAmu}
\end{eqnarray}
where the subscript $R$ and $I$ indicate respectively the real and imaginary part of field $\psi$. By substituting Eq.\eqref{peq:dpsi}
and \eqref{peq:dAmu} into the EOMs, Eq.\eqref{dphi}-\eqref{constrain}, one obtains the EOM for the perturbations
\begin{equation}\label{dphiR}
\begin{split}
\partial_{t} \partial_{z} \delta \psi_{R}&=\frac{1}{2}\big(-A_{x}^{s 2} \delta \psi_{R}-k^{2} \delta \psi_{R}-z \delta \psi_{R}+i k \delta A_{y} \psi_{I}^{s}-2 A_{x}^{s} \delta A_{x} \psi_{R}^{s}\\
&+\partial_{x}^{2} \delta \psi_{R}+\partial_{z}^{2} \delta \psi_{R}-z^{3} \partial_{z}^{2} \delta \psi_{R}+\delta \psi_{I} \partial_{x} A_{x}^{s}+\psi_{I}^{s} \partial_{x} \delta A_{x}\\
&+2 A_{x}^{s} \partial_{x} \delta \psi_{I}^{s}+2 \delta A_{x} \partial_{x} \psi_{I}^{s}-\delta \psi_{I} \partial_{z} A_{t}^{s}-\psi_{I}^{s} \partial_{z} \delta A_{t}-2 A_{t}^{s} \partial_{z} \delta\psi_{I}\\
&-3 z^{2} \partial_{z} \delta \psi_{R}-2 \delta A_{t} \partial_{z} \psi_{I}^{s}\big),
\end{split}
\end{equation}
\begin{equation}\label{dphiI}
\begin{split}
\partial_{t} \partial_{z} \delta \psi_{I}&=\frac{1}{2}(-A_{x}^{s 2} \delta \psi_{I}-k^{2} \delta \psi_{I}-z \delta \psi_{I}-2 A_{x}^{s} \delta A_{x} \psi_{I}^{s}-i k \delta A_{y} \psi_{R}^{s}\\
&+\partial_{x}^{2} \delta \psi_{I}+\partial_{z}^{2} \delta \psi_{I}-z^{3} \partial_{z}^{2} \delta \psi_{I}-\delta \psi_{R} \partial_{x} A_{x}^{s}-\psi_{R}^{s} \partial_{x} \delta A_{x}\\
&-2 A_{x}^{s} \partial_{x} \delta \psi_{R}-2 \delta A_{x} \partial_{x} \psi_{R}^{s}+\delta \psi_{R} \partial_{z} A_{t}^{s}+\psi_{R}^{s} \partial_{z} \delta A_{t}-3 z^{2} \partial_{z} \delta\psi_{I}\\
&+2 A_{t}^{s} \partial_{z} \delta \psi_{R}+2 \delta A_{t} \partial_{z} \psi_{R}^{s}),
\end{split}
\end{equation}
\begin{equation}\label{dAx}
\begin{split}
\partial_{t} \partial_{z} \delta A_{x}&=\frac{1}{2}\big(-k^{2} \delta A_{x}-4 A_{x}^{s} \delta \psi_{I} \psi_{I}^{s}-2 \delta A_{x} \psi_{I}^{s 2}-4 A_{x}^{s} \delta \psi_{R} \psi_{R}^{s}\\
&-2 \delta A_{x} \psi_{R}^{s 2}+\partial_{z}^{2} \delta A_{x}-z^{3} \partial_{z}^{2} \delta A_{x}-i k \partial_{x} \delta A_{y}+2 \psi_{R}^{s} \partial_{x} \delta \psi_{I}\\
&-2 \psi_{I}^{s} \partial_{x} \delta \psi_{R}+2 \delta \psi_{R} \partial_{x} \psi_{I}^{s}-2 \delta \psi_{I} \partial_{x} \psi_{R}^{s}-3 z^{2} \partial_{z} \delta A_{x}+\partial_{x}\partial_{z} \delta A_{t}\big),
\end{split}
\end{equation}
\begin{equation}\label{dAy}
\begin{split}
\partial_{t} \partial_{z} \delta A_{y}&=\frac{1}{2}\big(-2 i k \delta \psi_{R} \psi_{I}^{s}-2 \delta A_{y} \psi_{I}^{s 2}+2 i k \delta \psi_{I} \psi_{R}^{s}-2 \delta A_{y} \psi_{R}^{s 2}\\
&+\partial_{x}^{2} \delta A_{y}+\partial_{z}^{2} \delta A_{y}-z^{3} \partial_{z}^{2} \delta A_{y}-i k \partial_{x} \delta A_{x}+i k \partial_{z} \delta A_{t}-3 z^{2} \partial_{z} \delta A_{y}\big),
\end{split}
\end{equation}
\begin{equation}\label{dAt}
\begin{split}
\partial_{z} \partial_{t} \delta A_t&=-\big(\delta A_t k^2+4 A_t^{s} \delta \psi_I \psi_I^{s}+2 \delta{A_t}  \psi_{I}^{s 2}+4 A_t^{s} \delta \psi_R  \psi_R^{s}+2 \delta A_t \psi_{R}^{s 2}\\
&-\partial_{{x}}^2 \delta A_t+i k \partial_{t} \delta A_y-2 \psi_R^{s} \partial_{t} \delta \psi_I+2 \psi_I^{s} \partial_{t} \delta \psi_R-2 \delta \psi_R \partial_{t} \psi_I^{s}\\
&+2 \delta \psi_I \partial_{t} \psi_R^{s}-i k \partial_{z} \delta A_y+i k{z}^3 \partial_{z} \delta A_y+2 \psi_R^{s} \partial_{z} \delta \psi_I\\
&- 2 \psi_I^{s} \partial_{z} \delta \psi_R+2 z^3 \psi_I^{s} \partial_{z} \delta \psi_R+2 \delta \psi_R \partial_{z} \psi_I^{s}-2 \delta \psi_R z^3 \partial_{z} \psi_I^{s}\\
&- 2 \delta \psi_I \partial_{z} \psi_R^{s}+2 \delta \psi_I z^3 \partial_{z} \psi_R^{s}+\partial_{x} \partial_{t} \delta A_x-\partial_{x} \partial_{z} \delta A_x\\
&-2 z^3 \psi_R^{s} \partial_{z} \delta \psi_I+z^3 \partial_{x} \partial_{z} \delta A_x\big),
\end{split}
\end{equation}
and the constrain function is
\begin{equation}\label{d_constr}
\begin{split}
\partial_{z}^{2} \delta A_{t}&=i k \partial_{z} \delta A_{y}-2 \psi_{R}^{s} \partial_{z} \delta \psi_{I}+2 \psi_{I}^{s} \partial_{z} \delta \psi_{R}-2 \delta \psi_{R} \partial_{z} \psi_{I}^{s}+2 \delta \psi_{I} \partial_{z} \psi_{R}^{s}+\partial_{x} \partial_{z} \delta A_{x}.
\end{split}
\end{equation}
Together with the boundary conditions forthe  perturbation fields
\begin{equation}
\delta \psi_{R,I} |_{z=0}=0,\ \ \delta A_{\mu}|_{z=0}=0,\ \ \partial_z \delta A_t |_{z=0}=- \delta \rho, \label{p:bcs}
\end{equation}
one eventually obtains the matrix valued problem defined in the main text, which can be solved by Floquet analysis.
\subsection{Time crystal and synchronized phases in Fourier space}
\begin{figure}[h]
\begin{subfigure}{0.65\textwidth}
\includegraphics[width =\textwidth]{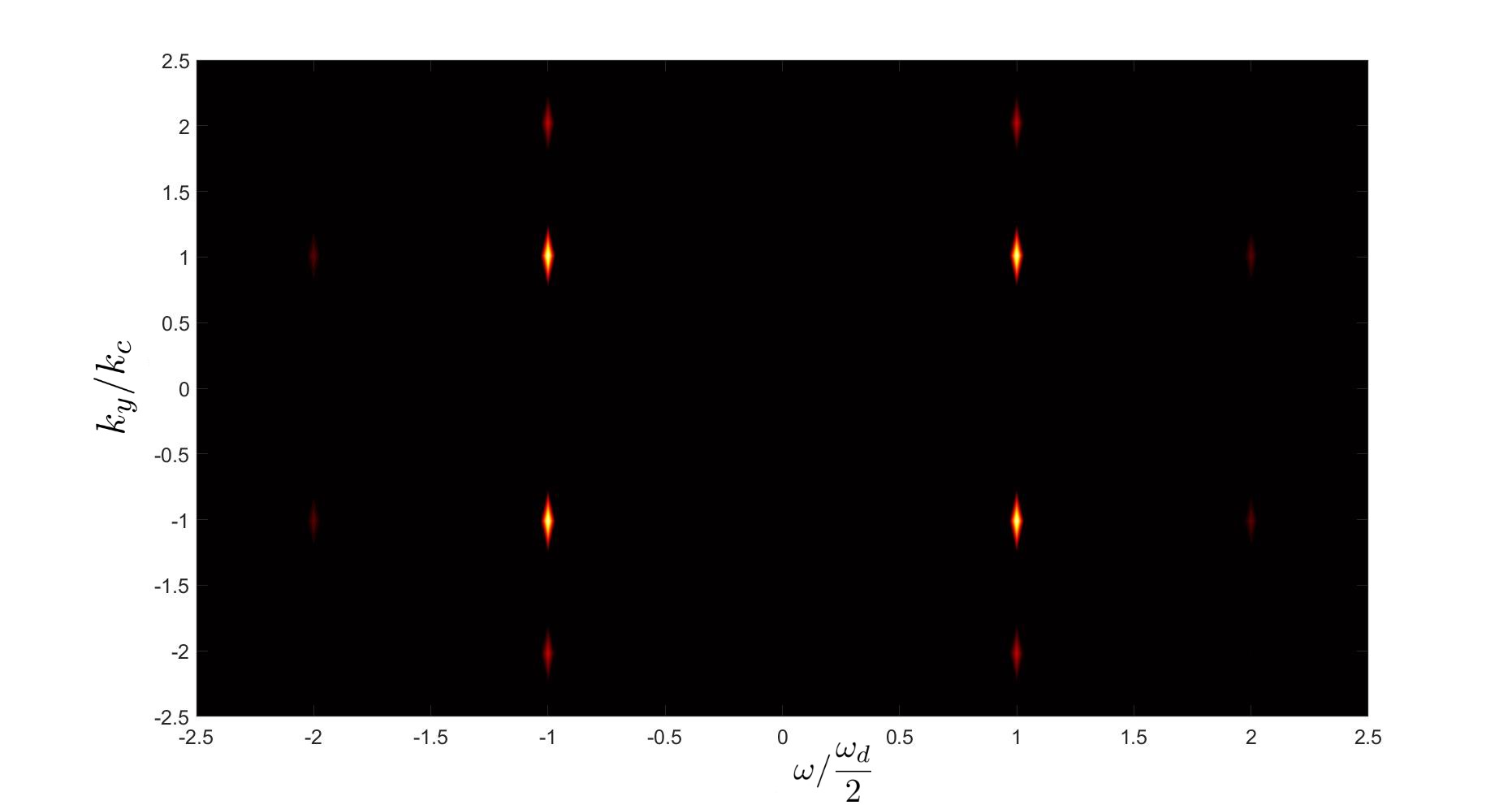}
\caption{}
\label{fig_tc_rho}
\end{subfigure}
\\
\begin{subfigure}{0.65\textwidth}
\includegraphics[width =\textwidth]{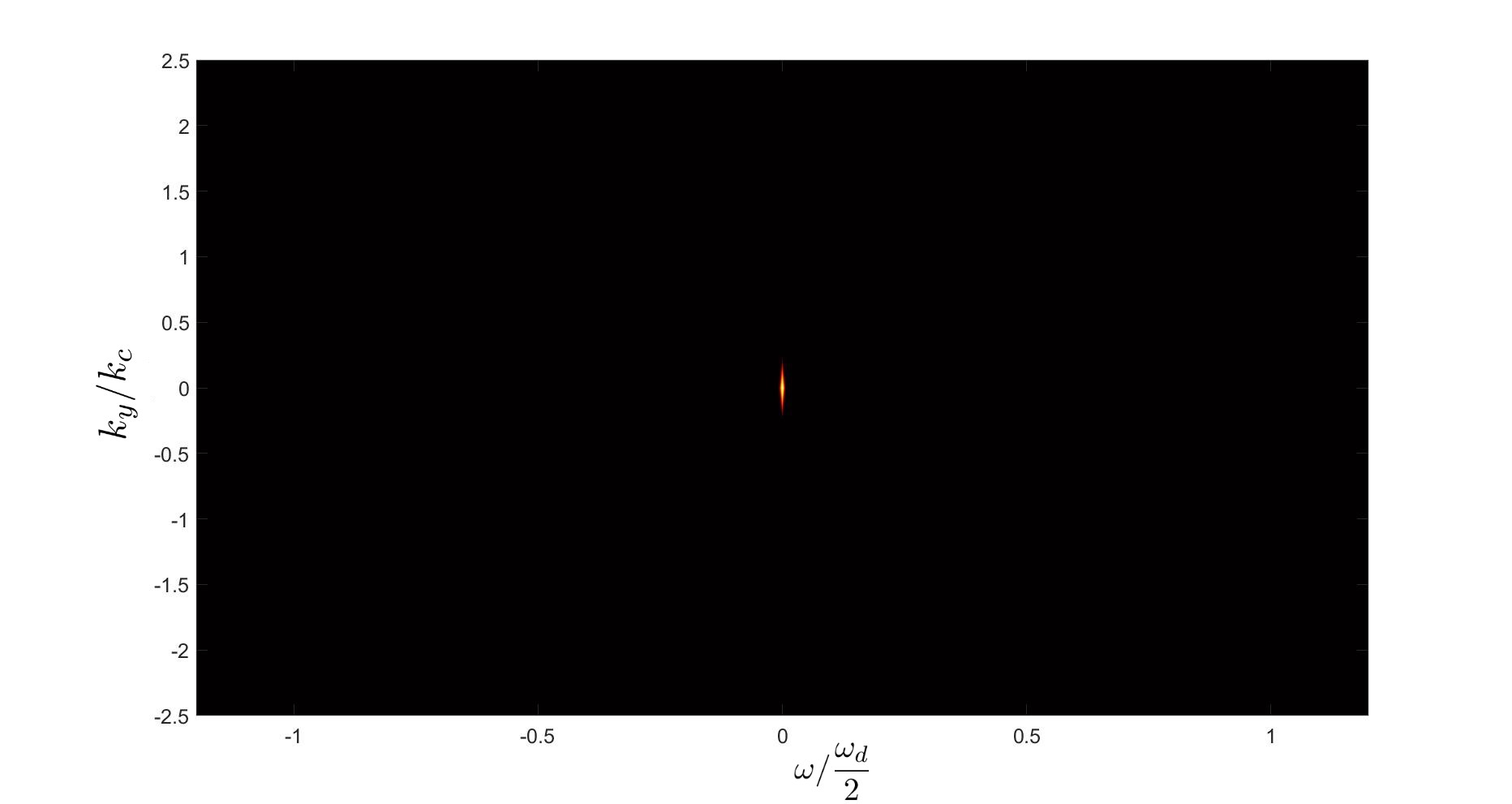}
\caption{}
\label{fig_syn_rho}
\end{subfigure}
\\
\begin{subfigure}{0.65\textwidth}
\includegraphics[width =\textwidth]{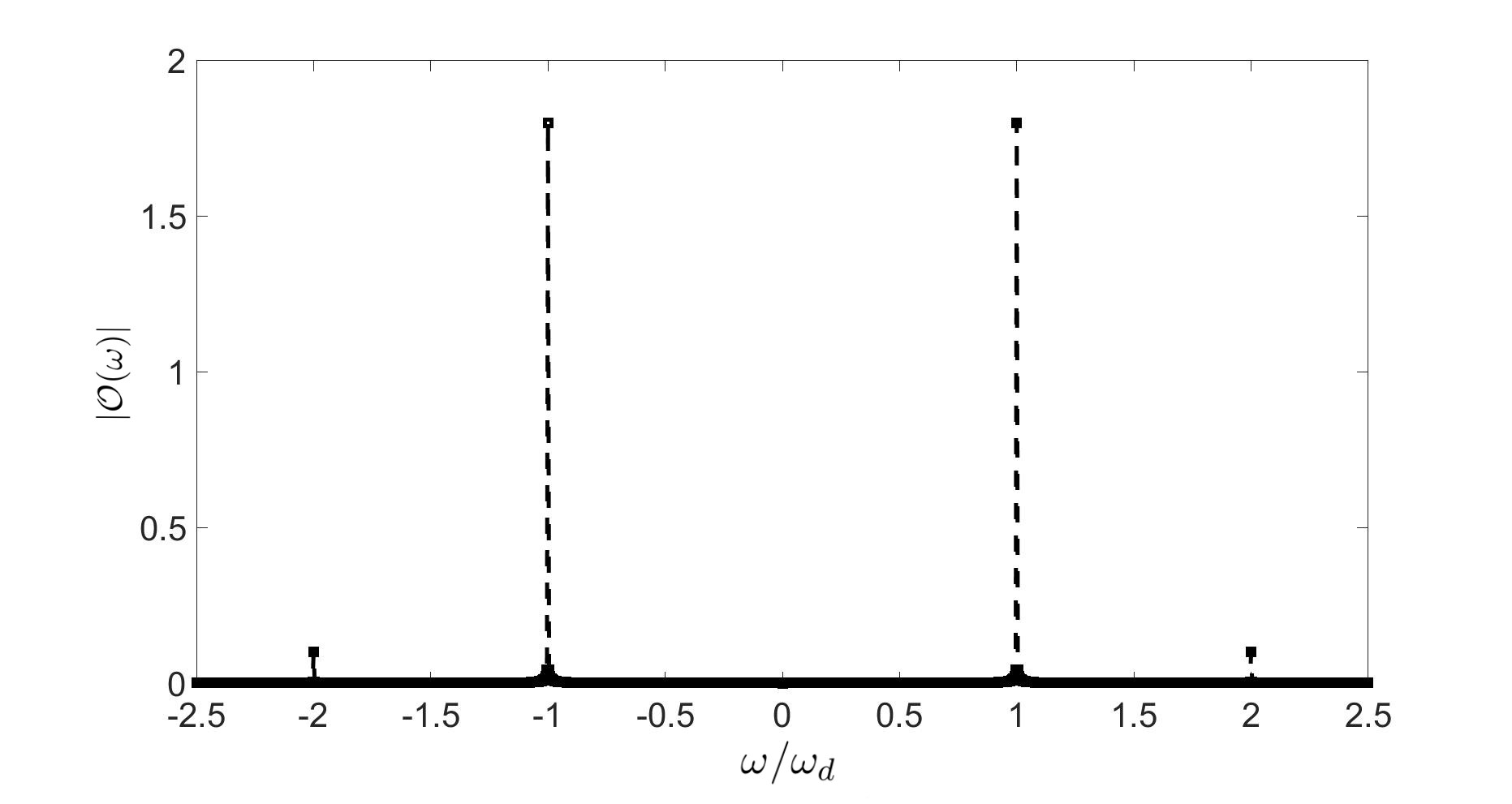}
\caption{}
\label{fig_syn_O}
\end{subfigure}
\caption{Fig.\ref{fig_tc_rho} is the Fourier spectrum $\rho(\omega, k_y)$ in the STS phase with $\tau=0.5$. The dominant mode locates at $(\pm k_c, \pm \frac{\omega_d}{2})$ and signals the breaking of time translations. Fig.\ref{fig_syn_rho} is the Fourier spectrum $\rho(\omega, k_y)$ of the synchronized SF phase with $\tau=0.7$. As expected, there is only a zero mode $(0,0)$ corresponding to the homogeneous and static $\rho_x(t,y)$.  Fig.\ref{fig_syn_O} is the Fourier spectrum $\mathcal{O}(\omega)$ of the synchronized SF phase with $\tau=0.7$. The dominant mode appears at $(\pm {\omega_d})$. All of these spectra are calculated with $\Theta=1.5 \omega_d$.}
\label{fig_fft}
\end {figure}

As mentioned in the main text, the space-time supersolid phase can be characterized by the Fourier modes $\tilde \rho=\rho(\pm k_c, \frac{\omega_d}{2})$ for the normalized average density $\rho_x(t,y)$. The space-time supersolid phase will have a non-zero contribution from $\tilde \rho$ (Fig.\ref{fig_tc_rho}), while the synchronized superfluid phase doesn't have any contribution from $\tilde \rho$ (Fig.\ref{fig_syn_rho}). To distinguish the synchronized superfluid phase from the normal fluid phase, one can check the Fourier modes $\mathcal{O}(\omega_d)$ of the superfluid order parameter $\langle \mathcal{O}(t,\mathbf{r}_0) \rangle$. Similarly, for synchronized superfluid phase we have $\mathcal{O}(\omega_d)\neq 0$ (Fig.\ref{fig_syn_O}), but for normal fluid phase we have $\mathcal{O}(\omega_d)=0$.

\newpage
\clearpage
\subsection{Heating rate with backreaction in a toy model}
In the main text, we have restricted our computations to the probe limit. As argued therein, this is equivalent to considering a dissipative open system coupled to a thermal bath with a parameterically large number of degrees of freedom. In this limit, the system can be kept in a steady state with constant energy and constant temperature for an infinite time.\\

Here, we want to show that even by introducing backreaction, the heating rate can be controlled by the backreaction strength $\kappa=16\pi G$ and made arbitrarily slow. In order to prove explicitly this fact, we resort to a simplified toy model. We consider the following action
\begin{equation}\label{eqp}
    S=\int d^4x \left[\frac{R-2\Lambda}{\kappa}-\left(\left(\partial \phi\right)^2+m^2 \phi^2\right)\right],
\end{equation}
where $\kappa$ determines the backreaction strength through the Einstein equation
\begin{equation}
    G_{ab}+\Lambda g_{ab}=\frac{1}{2}\kappa T^\phi_{ab}\,,\label{eqf2}
\end{equation}
with $T^\phi_{ab}$ the scalar bulk stress-energy tensor.
We take a time-dependent ansatz for the metric given by
\begin{equation}\label{Eddington}
    ds^{2}=\frac{L^{2}}{z^{2}}\left(-f(t,z)e^{-\chi(t,z)}dt^{2}-2e^{-\chi(t,z)}dt dz+dx^{2}+dy^{2}\right)
\end{equation}
with the AdS radius $L=1$. In addition, we choose the mass of the scalar $m^2=-2$ such that its UV behavior is given by
\begin{equation}
    \phi=\phi_1(t)z+\phi_2(t)z^2+\cdots\,.
\end{equation}
Accordingly, the asymptotic behavior of the metric can be obtained as follows
\begin{eqnarray}
    f&=&1+\kappa\frac{\phi_1(t)^2}{4}z^2-\left[E(t)-\frac{\kappa}{3}\phi_1(t)\phi_2(t)\right]z^3+\cdots,\\
    \chi&=&\kappa\left(\frac{\phi_1(t)^2}{4}z^2+\frac{2}{3}\phi_1(t)\phi_2(t)z^3+\cdots\right)
\end{eqnarray}
with $E(t)$ the time-dependent energy density of the boundary field theory by the standard holographic dictionary.
We are interested in a setup in which the scalar operator dual to the bulk scalar $\phi$ is driven by a time-dependent source
\begin{equation}
    \phi_1(t)=\mathcal{A}\sin (\omega_D t).
\end{equation}
We study the increase of energy density as a function of time, \textit{i.e.}, the heating rate, for different values of the backreaction parameter $\kappa$. With these notations, $\kappa=0$ corresponds to the probe limit presented in the main text, in which the energy density remains constant in time.
\begin{figure}[ht!]
    \centering\includegraphics[width=0.5\linewidth]{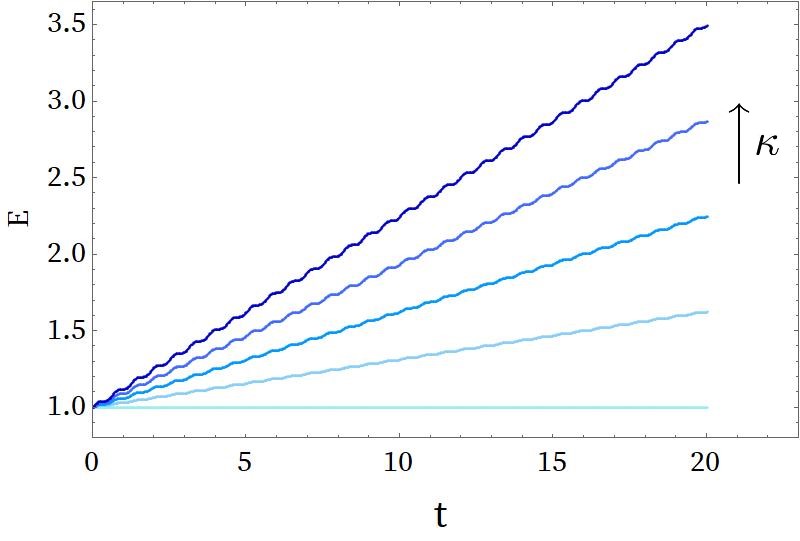}
    \caption{The growth of the dual field theory energy as a function of time in the scalar toy model, setting $\mathcal{A}=0.1$ and $\omega_D=5.0$. The lines from bottom to top correspond respectively to $\kappa=0,0.25,0.5,0.75,1$.}
    \label{fig:f}
\end{figure}
In Fig.\ref{fig:f}, we show the increase of the energy density for constant amplitude and driving frequency but for different values of $\kappa$. As one can see, $\kappa$ controls the rate of increase of the energy density upon driving the system. This rate can be made arbitrarily small by tuning $\kappa$. From a field theory perspective, this is a direct consequence of the Ward identity 
\begin{equation}
    \partial_t E= \kappa \langle \mathcal{O}\rangle\partial_t\phi_1,\label{eqf3}
\end{equation}
with $\langle \mathcal{O}\rangle=\partial_t\phi_1-\phi_2$ (due to the Eddington-Finkelstein-like coordinates (\ref{Eddington}) used here) the vacuum expectation value, whereby the heating rate is proportional to $\kappa$. 

This simple toy model shows that even in presence of backreaction the heating rate can be controlled and made arbitrarily slow by tuning down $\kappa$, corresponding to the dual boundary system immersed in a well behaved thermal reservoir with the parametrically large number of degrees of freedom characterized by $O(N^2)\propto\frac{1}{\kappa}$.
\end{document}